\documentclass[prl,showpacs,twocolumn]{revtex4}

\newcommand {\eg}{$e_{\mathrm{g}}$}
\newcommand {\myub}{$\mu_{\mathrm {B}}$}
\newcommand {\Jone}{$J_1$}
\newcommand {\Jtwo}{$J_2$}
\newcommand {\Jdc}{$J_{\mathrm {DC}}$}
\newcommand {\SdotS}{{\bf S}_i \cdot {\bf S}_j}
\newcommand {\StimesS}{{\bf S}_i \times {\bf S}_j}
\newcommand {\Cutwo}{Cu$^{2+}$}
\newcommand {\Cuone}{Cu$^{1+}$}
\newcommand {\CFO}{CuFeO$_2$}
\newcommand {\TMO}{TbMnO$_3$}
\newcommand {\LCO}{LiCu$_2$O$_2$}
\newcommand {\cutwo}{Cu$^{2+}$}
\newcommand {\Sn }{${\bf S}_{\mathrm{n}}$}
\newcommand {\Qvec}{${\bf Q}$}
\newcommand {\SperpQ}{\Sn $\bot$\Qvec}
\newcommand {\SparaQ}{\Sn $\| $\Qvec}
\newcommand {\kscan}{$k$-scan}

\newcommand {\Tnone}{$T_{\mathrm{N1}}$}
\newcommand {\Tntwo}{$T_{\mathrm{N2}}$}
\newcommand {\etal}{{\it et al.}}
\newcommand {\chiT}{d$\chi$/d$T$}
\newcommand {\ki}{{\bf k}$_{\mathrm {i}}$}
\newcommand {\kf}{{\bf k}$_{\mathrm {f}}$}
\newcommand {\ea}{{\bf e}_a}
\newcommand {\eb}{{\bf e}_b}
\newcommand {\ec}{{\bf e}_c}
\newcommand {\normalscat}{\Bigl( \frac{d\sigma }{d\Omega} \Bigr )}
\newcommand {\smallscat}{( {d\sigma }/{d\Omega})}

\newcommand {\qhat}{\hat{{\bf Q}}}

\newcommand {\qdotChirality}{(\qhat \cdot {\bf C})}
\newcommand {\qm}{{\bf q}_m}
\newcommand {\mb}{m_b}
\newcommand {\mc}{m_c}
\newcommand {\ma}{m_a}
\newcommand {\mi}{{\bf m}_i}

\newcommand {\Chirality}{{\bf C}}
\newcommand {\qvec}{{\bf Q}}
\newcommand {\magmag}{{\bf \eta}}
\newcommand {\Etai}{\magmag_i}
\newcommand {\Etaj}{\magmag_j}
\newcommand {\Sigmaix}{{\bf \sigma}_i^x}
\newcommand {\Sigmaiy}{{\bf \sigma}_i^y}

\newcommand {\Sigmapm}{{\bf \sigma}^\pm}
\newcommand {\Sigmax}{{\bf \sigma}^x}

\newcommand {\Sigmay}{{\bf \sigma}^y}

\newcommand {\Sigmaip}{{\bf \sigma}_i^+}
\newcommand {\Sigmaim}{{\bf \sigma}_i^-}
\newcommand {\Sigmajp}{{\bf \sigma}_j^+}
\newcommand {\Sigmajm}{{\bf \sigma}_j^-}

\newcommand {\Ion}{I_{\mathrm {ON}}}
\newcommand {\Ioff}{I_{\mathrm {OFF}}}
\newcommand {\HatSn}{\hat{\bf S}_{\mathrm{n}}}
\newcommand {\SSn }{{\bf S}_{\mathrm{n}}}

\usepackage{graphicx}
\usepackage{bm}
\usepackage{pifont}
\usepackage{amsmath}

\begin{document}

\title{Correlation between spin helicity and electric polarization vector in quantum-spin chain magnet LiCu$_2$O$_2$}

\author{S. Seki$^1$, Y. Yamasaki$^1$, M. Soda$^2$, M. Matsuura$^2$, K. Hirota$^2$ and Y. Tokura$^{1,3}$} 
\affiliation{$^1$ Department of Applied Physics, University of Tokyo, Tokyo 113-8656, Japan \\ $^2$ The Institute for Solid State Physics, University of Tokyo, Kashiwa 277-8581, Japan \\ $^3$  Multiferroics Project, ERATO, Japan Science and Technology Agency (JST), Tokyo 113-8656, Japan}

\date{}

\begin{abstract}

Measurements of polarized neutron scattering were performed on a $S=1/2$ chain multiferroic LiCu$_2$O$_2$. In the ferroelectric ground state with the spontaneous polarization along the $c$-axis, the existence of transverse spiral spin component in the $bc$-plane was confirmed. When the direction of electric polarization is reversed, the vector spin chirality as defined by ${\bf C}_{ij} = {\bf S}_i \times {\bf S}_j$ ($i$ and $j$ being the neighboring spin sites) is observed to be reversed, indicating that the spin-current model or the inverse Dzyaloshinskii-Moriya mechanism is applicable even to this $e_{\mathrm{g}}$-electron quantum-spin system. Differential scattering intensity of polarized neutrons shows a large discrepancy from that expected for the classical-spin $bc$-cycloidal structure, implying the effect of large quantum fluctuation.

\end{abstract}
\maketitle

\begin{figure}
\begin{center}
\includegraphics*[width=8cm]{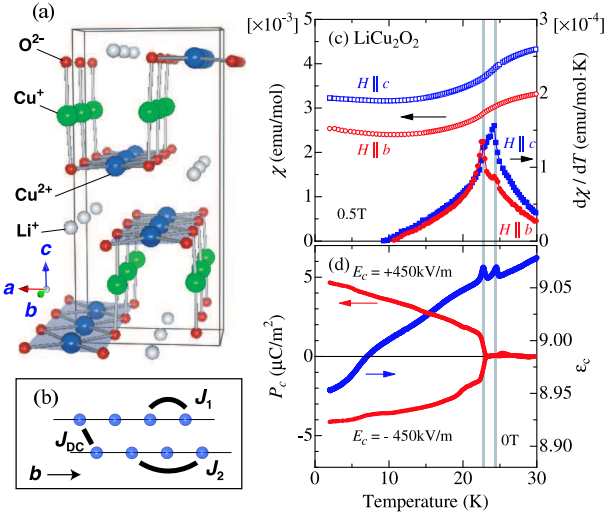}
\caption{(color online). (a)Crystal structure of {\LCO}. (b) Schematic view of magnetic interactions between {\cutwo} sites. (c),(d) Temperature dependence of magnetic susceptibility, electric polarization and dielectric constant. All the quantities were measured in the warming process.}
\end{center}
\end{figure}

Magnetoelectric effect, controlling dielectric (magnetic) properties by magnetic (electric) field, has been studied for long because of its potential for novel physics and application. Although several low-symmetric materials like Cr$_2$O$_3$ were found to show the linear magnetoelectric response, the effect has been very small\cite{Review1}. Recently, the phenomenon of electric polarization flop with magnetic field was found for perovskite type {\TMO}\cite{Kimura}. In this material, the specific magnetic structure itself induces ferroelectricity, which enables the colossal magnetoelectric responses via the magnetic phase transition\cite{Review2}. The key issue is the coupling mechanism between the spin habit and the polarization. Thus far, mainly two models have been proposed; the model based on the symmetric exchange interaction (${\SdotS}$) or the antisymmetric one (${\StimesS}$)\cite{Review3}. For the latter case, a microscopic model was devised by Katsura, Nagaosa and Balatsky (KNB)\cite{Katsura}, in which the electric polarization ${\bf P}_{ij}$ produced between magnetic moments at neighboring sites $i$ and $j$ (${\bf m}_{i}$ and ${\bf m}_{j}$) is given as
\begin{equation}
\label{KatsuraFormula}
{{\bf P}_{ij}} = A \cdot {\bf e}_{ij} \times ({\bf m}_{i} \times {\bf m}_{j})
\end{equation}
Here, ${\bf e}_{ij}$ is the unit vector connecting the site $i$ and $j$, and $A$ a coupling constant related to the spin-orbit and exchange interactions. This model predicts that a helimagnet with transverse spiral components can be ferroelectric, and well explains the ferroelectric behaviors observed for $R$MnO$_3$ ($R$ = Tb and Dy) \cite{Kimura, Kimura2, Goto}, Ni$_3$V$_2$O$_8$\cite{Ni3V2O8}, CoCr$_{2}$O$_4$\cite{CoCr2O4}, MnWO$_4$\cite{MnWO4}, etc. All these materials contain the frustration of magnetic interactions as a source of noncollinear spin structure. Besides, the ferroelectric spiral magnets based on other than KNB model, such as {\CFO}\cite{CuFeO2,Arima,Nakajima,seki}, have also been reported. These materials with both magnetic and dielectric orders are now broadly termed multiferroics.

{\LCO}, as investigated here, has recently been found to be one such member of multiferroics\cite{Cheong1}. Figure 1(a) indicates the crystal structure of {\LCO}; the space group $Pnma$, and lattice parameters $a = 5.73$, $b= 2.86$ and $c = 12.4${\AA} at room temperature\cite{Masuda}. This material contains equal number of {\Cuone} and {\Cutwo}, only the latter of which carries spin $S$=1/2. Each {\Cutwo} ion is on the center of oxygen square and forms edge-shared chains running along the $b$-axis with the Cu-O-Cu bond angle of 94$^\circ$. As expected from the Kanamori-Goodenough rule, the nearest neighbor exchange interaction ({\Jone}) is ferromagnetic though relatively weak as compared with the antiferromagnetic next nearest neighbor interaction ({\Jtwo}), causing the magnetic frustration. The magnitude of inter-chain interaction ({\Jdc}) is presumed to be small($<|J_1|,|J_2|$), though has not reached the consensus as yet (Fig. 1 (b))\cite{Gippius,Masuda2}. As a result of the frustration, a spiral magnetic structure is realized below {\Tntwo}$\sim$ 23K. A former (unpolarized) neutron diffraction study has revealed the incommensurate magnetic structure with the modulation vector (0.5, 0.174, 0), and claimed the $ab$-spiral state\cite{Masuda}. In this phase, however, the appearance of spontaneous electric polarization along the $c$-axis has recently been reported\cite{Cheong1}. To reconcile the observed polarization direction with the spiral spin state, the KNB model requires the $bc$-spiral spin structure. Recent resonant soft x-ray magnetic scattering study suggests a more complex spin spiral\cite{Cheong2}, and the magnetic structure of the ferroelectric ground state is still under controversy. Incidentally, the powder neutron study on the isostructural material NaCu$_2$O$_2$ justifies the $bc$-spiral spin structure, while the magnetic moment of {\Cutwo} is estimated as small as 0.56{\myub}\cite{Keimer}. This implies that the effect of quantum fluctuation is important also in {\LCO}. 

In this paper, to clarify the origin of ferroelectricity in {\LCO}, we testify the validity of the KNB model for the {\eg}-electron spin system with potentially large quantum fluctuation. Recently, the polarized neutron scattering experiment on {\TMO} has confirmed the coupling between the spin vector chirality and the direction of electric polarization in accord with the KNB model\cite{Yamasaki}. Since the polarity-dependent vector chirality can be the definitive evidence for the spiral-spin driven ferroelectricity, we performed the related experiments on {\LCO}.

Single crystals of {\LCO} were grown by the self-flux method. Under a polarized optical microscope, the fine twin structure with mixing of the $a$ and $b$-axis domains was observed in accord with the former observations \cite{Cheong1,twin}. The crystal was cleaved into a thin plate with the widest faces parallel to (001) plane. As the electrodes, Al was deposited on the $ab$ faces. Polarized neutron diffraction experiments were carried out with the ISSP-PONTA triple-axis spectrometer at JRR-3M using a Heusler polarizer. In this paper, we define the scattering vector {\bf Q} as {\bf Q} = {\kf} - {\ki}, where {\ki} and {\kf} are the wave vectors of the incident and diffracted neutrons, respectively. The polarization direction of incident neutrons ({\Sn}), as defined by the magnetic field ($\sim10$ mT) generated with a Helmholtz coil, can be reversed by a neutron-spin flipper. The polarized neutron scattering experiments were executed for the two configurations; {\SperpQ} and {\SparaQ} (Figs. 2 (a) and (b)). The flipping ratio of polarized to unpolarized neutrons measured at the (2,1,0) nuclear reflection was sufficiently large; 33 for {\SperpQ} and 27 for {\SparaQ}. The sample was mounted on a sapphire plate in a closed-cycle helium refrigerator, so that the horizontal scattering plane of the spectrometer coincided with the ($h$ $k$ 0) zone. The neutron energy was fixed at 13.47 meV and the collimations $40'-40'-40'-80'$ were employed. Higher-order neutrons were removed by a pyrolytic graphite filter. The size of the specimen used for the neutron study is 12 mm$^2$ ($ab$ plane) $\times 0.6$ mm ($c$-axis). All the data presented in this paper were measured on the identical sample. Dielectric constant was measured at 100kHz using an $LCR$ meter. For the electric polarization, we measured the pyroelectric current with a constant rate of temperature sweep ($\sim$2K/min) and integrated it with time. To obtain a single ferroelectric domain, the poling electric field was applied in the cooling process and removed just before the measurements of pyroelectric current and polarized neutron scattering. Magnetization was measured with a Magnetic Property Measurement System (Quantum Design Inc.).

Figures 1 (c) and (d) show the temperature dependence of magnetic susceptibility, dielectric constant, and electric polarization for {\LCO}. For $H\parallel c$, the temperature derivative of magnetic susceptibility ({\chiT}) indicates two anomalies at {\Tnone}$\sim$24.5K and {\Tntwo}$\sim$23.0K, although only one peak at {\Tntwo} is found in {\chiT} for $H\parallel b$ (or $a$). These imply the existence of two magnetic phases below {\Tnone}; AF1 ({\Tnone}$> T >${\Tntwo}) and AF2 ({\Tntwo}$> T$). The anomaly at 9K possibly caused by  impurity Li$_2$CuO$_2$\cite{Masuda,twin} was absent in our sample. The spontaneous electric polarization parallel to the $c$-axis ($P_c$) evolves only below {\Tntwo}. The $P_c$ can be reversed with the opposite poling electric field ($E_c$). This indicates the ferroelectric nature of AF2 phase, and suggests the correlation between ferroelectricity and magnetic properties. All these features reproduced the results reported by S. Park {\it et al} \cite{Cheong1,Cheong2}, who proposed the sinusoidal spin structure with collinear spins (parallel to the $c$-axis) for AF1. A recent theory proposed the intriguing scenario of the novel cholesteric spin state for this phase\cite{Onoda2}. We measured the poling electric field dependence of spontaneous polarization and confirmed that the saturation of $P_c$ was achieved above $|E_c| \sim$ 350kV/m. We also measured dielectric constant parallel to the $c$-axis ($\epsilon _c$) and found peaks at both {\Tnone} and {\Tntwo}, although previously only one peak at {\Tntwo} was reported\cite{Cheong1}. 

\begin{figure}
\begin{center}
\includegraphics*[width=8cm]{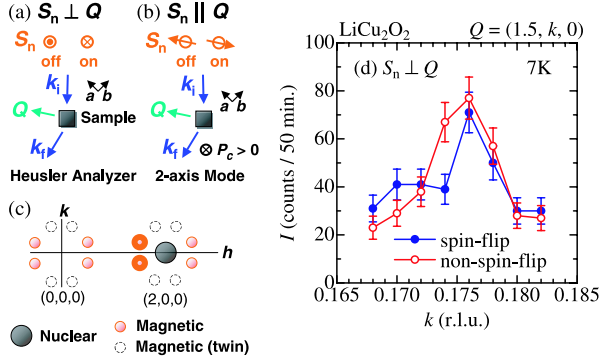}
\caption{(color online). The experimental geometries for the polarized neutron diffraction; (a) {\SperpQ} and (b) {\SparaQ}. The labels ``on'' and ``off'' indicate the state of the neutron-spin flipper. (c) Schematic illustration of nuclear and magnetic Bragg positions in the reciprocal space. (d) The {\kscan} profiles of the (1.5, +$\delta$, 0) magnetic reflection in the {\SperpQ} setup. }
\end{center}
\end{figure}

For the polarized neutron diffraction measurements, we focused on the ferroelectric AF2 phase. Since different magnetic structures, such as the $ab$-spiral\cite{Masuda} and the $bc$-spiral plus $a$-component structure\cite{Cheong1,Cheong2}, have been proposed for this phase, whether the magnetic moment is present along the $c$-axis was first examined. For this purpose, we took the {\SperpQ} setup (Fig. 2(a)), where neutron spins were parallel or antiparallel to the $c$-axis. To distinguish between the spin-flip and non-spin-flip scattering, a Heusler analyzer was employed. In general, only the magnetic moment perpendicular to {\bf Q} contributes to the magnetic reflection of neutrons. For polarized neutrons, furthermore, the magnetic moment parallel to {\Sn} produces the non-spin-flip scattering and the moment perpendicular to {\Sn} does the spin-flip scattering\cite{Moon}. Figure 2 (d) shows the {\kscan} profile of the (1.5, +$\delta$, 0) magnetic reflection at 7K ($<$ {\Tntwo}). The observed modulation wavenumber, $\delta\sim0.175$, is in accord with literature\cite{Masuda}. Since {\bf Q} can be considered almost parallel to the $a$-axis in this configuration (Fig. 2(c)), the $b$-component of magnetic moment ($m_b$) contributes to the spin-flip scattering while the $c$-component ($m_c$) to the non-spin-flip scattering. Assuming the common background for the both profiles, the integrated intensities are nearly equal (spin-flip($m_b$) / non-spin-flip($m_c$) $\approx 0.9$). This suggests the existence of the nearly same weight of $b$- and $c$-components in the magnetic structure of AF2. This is consistent with the $bc$-spiral (or plus some $a$-component) model\cite{Cheong1,Cheong2}, and at least not with the simple $ab$-spiral one\cite{Masuda}.

Next, we attempted to observe the relationship between the polarization direction and the chirality of spin spiral. For this purpose, we adopted the {\SparaQ} setup (Fig. 2(b)), where neutron spins are parallel or antiparallel to {\bf Q}. In this alignment, only spin-flip scatterings contribute to the magnetic reflection. Therefore, no polarization analysis is needed, and we employed the two-axes mode without an analyzer. Figures 3 (a)-(d) show the $k$-scan profiles of the (1.5, $\pm $$\delta$, 0) magnetic reflection at 7K\cite{Note3} with various poling electric fields parallel to the $c$-axis ($E_c$). $E_c$ was applied at 30K ($>$\Tnone) and removed at 7K just before the diffraction measurements to obtain a single ferroelectric domain. With $|E_c|= 450$kV/m, the difference of intensity between $\pm \delta$ was clearly observed, and the relative intensity was confirmed to be reversed by changing the sign of either {\Sn} or $E_c$. These behaviors can be interpreted in terms of the $E_c$-dependent vector chirality of the transverse $bc$-spiral spins as follows. 

\begin{figure}
\begin{center}
\includegraphics*[width=8cm]{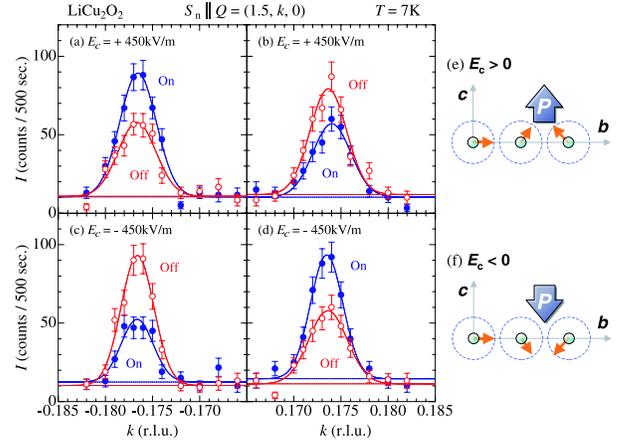}
\caption{(color online). (a)-(d) The {\kscan} profiles of the (1.5, $\pm $$\delta$, 0) magnetic reflections in the {\SparaQ} setup. The labels ``on'' and ``off'' show the state of neutron-spin flipper. Solid lines show the result of the Gaussian fitting. (e), (f) The geometrical relationships between spin chirality (helicity) and electric polarization determined from the observed results.}
\end{center}
\end{figure}

According to Blume\cite{Blume}, the magnetic cross section for polarized neutron is given as
\begin{equation}
\label{BlumeCross}
\begin{split}
\normalscat \propto \sum_{i,j}\exp \{ i{\bf Q}({\bf R}_i-{\bf R}_j) \} \bigl[ \Etaj \cdot \Etai + i \HatSn (\Etaj \times \Etai) \bigr]
\end{split}
\end{equation}
Here, $\Etai$  denotes the component of $\mi$ perpendicular to {\bf Q}, $\Etai=\qhat \times (\mi \times \qhat)$, where $\qhat = {\bf Q}/|{\bf Q}|$ and $\HatSn = {\SSn}/|{\SSn}|$. For simplicity, we take hereafter the approximation that {\Sn}$\parallel \qvec \parallel a$ and define $abc\rightarrow zxy$, where $z$ is the spin quantization axis. Then, the spin vector chirality on the $bc$-plane can be defined as $\Chirality = (\Etai \times \Etaj) / |\Etai \times \Etaj|$. With use of the relations $\Etai=(\Sigmaix,\Sigmaiy,0)$ and $\Sigmapm=\Sigmax \pm i\Sigmay$, the cross section for the $(1.5, \pm \delta, 0)$ magnetic reflections can be expressed as
\begin{equation}
\label{QuantumCrossSection}
\begin{split}
\normalscat_\pm=\normalscat_c \pm \normalscat_s
\end{split}
\end{equation}
where
\begin{equation}
\label{Quantum1}
\begin{split}
\normalscat_c \propto \sum_{i,j}\cos \{{\bf Q}({\bf R}_i-{\bf R}_j) \} \cdot \langle \Sigmaip \Sigmajm \rangle
\end{split}
\end{equation}
\begin{equation}
\label{Quantum2}
\begin{split}
\normalscat_s \propto \sum_{i,j}\sin \{{\bf Q}({\bf R}_i-{\bf R}_j) \} \cdot \langle \Sigmaip \Sigmajm \rangle
\end{split}
\end{equation}
For intuitive understanding, we tentatively treat the cross section in the classical limit. Based on the results for the {\SperpQ} setup, we can assume the $bc$-spiral magnetic structure plus several $a$-component: 
\begin{equation}
\label{MagneticStructure}
\begin{split}
{\bf m}_i = & \mb \cdot \eb \cdot \cos (\qm {\bf R}_i) + \mc \cdot \ec \cdot \sin (\qm {\bf R}_i)  \\ & + \ma \cdot \ea \cdot \sin (\qm {\bf R}_i + \delta^{'})
\end{split}
\end{equation}
Here, ${\ea}$, ${\eb}$, and ${\ec}$ are the unit vectors along the $a$, $b$, and $c$-axis. Then, Eq. (\ref{QuantumCrossSection}) can be written as \cite{Blume,Yamasaki}
\begin{equation}
\label{ClassicCross}
\begin{split}
\normalscat_\pm \propto \Bigl[ \mb^2 + \mc^2  \pm 2\mb \cdot \mc \cdot (\HatSn \cdot \qhat) \qdotChirality \Bigr]
\end{split}
\end{equation}
The last term predicts the different scattering intensities for $\pm \delta$, and the relation can be reversed by changing the sign of either {\Sn} or {\bf C}. In fact, this behavior is clearly observed in the results with $E_c$=+450kV/m (Figs. 3 (a) and (b)). This means that $\qhat \cdot{\bf C}$ is not zero, or in other words the magnetic structure of AF2 has the spiral components in the $bc$-plane. Moreover, when the sign of $E_c$ is reversed, the differential intensity relation is also reversed (Figs. 3 (c) and (d)). This indicates that the spin chirality determines the direction of electric polarization. Conversely, the observed electric control of spin helicity directly proves that the ferroelectricity of {\LCO} originates from the transverse-spiral (cycloidal) spin structure. Thus, the KNB model holds good even for the {\eg}-electron spin system, or under possibly large quantum fluctuation inherent to the frustrated $S$=1/2 spins. The obtained geometric relation between spin chirality and electric polarization is illustrated in Figs. 3 (e) and (f). The sign of the coupling constant in Eq. (\ref{KatsuraFormula}) is negative ($A<0$), which agrees with the theoretical prediction \cite{Onoda}. Note that the sign of $A$\cite{Note4} is different from the case of {\TMO}\cite{Yamasaki}. We also measured the profiles with $E_c=0$ and found no difference for the intensity between $\pm \delta$ reflections nor between the neutron spin states. This should be due to the coexistence of opposite ferroelectric domains (or clockwise/counter-clockwise spin-spiral domains) for the zero electric-field case.

An unresolved problem at this stage is the ratio of scattering intensity between the stronger and weaker reflections. From Eq. (\ref{ClassicCross}), the elliptic ratio of the spiral spin, $\mb/\mc$ (or $\mc/\mb$), is estimated as $|({\sqrt{\Ion}-\sqrt{\Ioff}})/({\sqrt{\Ion}+\sqrt{\Ioff}})|$ for the case of classical spin\cite{Yamasaki}. On the basis of the data shown in Figs. 3 (a)-(d), this expression gives $m_c/m_b$ (or $m_b/m_c)=0.09 \sim 0.20$. On the other hand, the aforementioned results on the {\SperpQ} setup suggests the nearly equal value for $\mb$ and $\mc$. As the origin of this discrepancy, the coexistence of different polarity domains might be suspected. However, we confirmed the saturation of electric polarization at $|E_c|=350$kV/m, with the same (Al) electrode used in the neutron scattering study. Also on the same sample, the Ag electrode was tested to confirm the identical saturation value of electric polarization. Therefore, we believe that the single domain state was realized in the {\SparaQ} setup, and the above apparent discrepancy should be ascribed to a more intrinsic origin. The measured temperature (7K) might not be low enough to saturate the spin order. However, the $P$ value at 7K already reaches 80-90$\%$ of the 2K value (see Fig. 1 (d)); thermal fluctuation alone is not enough to decrease the spin ellipticity $\mc/\mb$. One of other possibilities is the effect of quantum fluctuation. In case of $S$=1/2 quantum-spin systems like {\LCO}, the validity of the classical-spin treatment as done in Eqs. (\ref{MagneticStructure}) and (\ref{ClassicCross}) is no longer guaranteed. For a more rigorous argument, we have to go back to Eqs. (\ref{QuantumCrossSection}) - (\ref{Quantum2}). According to these expressions, both $\smallscat_c$ and $\smallscat_s$ are the Fourier components (symmetric and antisymmetric, respectively) of the same physical quantity $\Sigmaip\Sigmajm$. Therefore, the distribution of scattering intensities reflects the balance between symmetric and antisymmetric components of $\Sigmaip\Sigmajm$ for the $S$=1/2 case. This may be the cause of the deviation from the Eq. (\ref{ClassicCross}). For example, in the extreme case of quantum fluctuation where the spins form the singlet state, the commutation that $\langle \Sigmaip \Sigmajm \rangle = \langle \Sigmajp \Sigmaim \rangle$ holds, therefore $\smallscat_s=0$ and no differential intensity should be observed. The experimental observation of shrunk magnetic moment \cite{Keimer} implies the large quantum fluctuation subsisting in the ordered spiral state. Therefore, the quantum fluctuation of the vector spin chirality is likely to result in the reduced differential $\pm \delta$ reflection intensity of polarized neutrons, as observed. In addition, several groups have implied that the magnetic structure of AF2 would be more complicated than the simple $bc$-spiral\cite{Cheong1,Cheong2}. This may also require some modification in Eq. (\ref{ClassicCross}). Note however that even with any other magnetic structure the observed difference for the opposite neutron spins $\SSn$ reflects the chirality in the $bc$-plane (see Eq. (\ref{BlumeCross})). For the thorough understanding, further analysis of the magnetic structure and its quantum dynamics will be needed.

In summary, the polarized neutron study was performed on the quantum-spin chain magnet {\LCO}. We confirmed the coupling between spin vector chirality of the transverse $bc$-spiral structure and the direction of electric polarization along the $c$-axis. This proves that even with the {\eg}-electron system under the large quantum fluctuation the spin-current model or the inverse Dzyaloshinskii-Moriya mechanism still works. The differential intensity of polarized neutron reflections show a clear deviation from that expected for the classical $bc$-spiral spin structure, implying the importance of quantum fluctuation in this $S=1/2$ helimagnet.

The authors thank T. Arima, N. Furukawa, N. Nagaosa, S. Onoda, H. Katsura, J. Fujioka, Y. Shimada, H. Sakai , S. Iguchi and Y. Onose for enlightening discussions. This work was partly supported by Grants-In-Aid for Scientific Research (Grant No. 16076205, 17340104, 19052002) from the MEXT of Japan.

\end{document}